\documentclass[showpacs,preprintnumbers,amsmath,amssymb,preprint]{revtex4}
\begin{document}

\title{\Large\bf $Z_3$ orbifold construction of
 $SU(3)^3$ GUT with $\sin^2\theta_W^0=\frac38$}
\author{Jihn E.
Kim\footnote{jekim@th.phyik.uni-bonn.de,\ \
jekim@phyp.snu.ac.kr}}
\address{
School of Physics, Seoul National University, Seoul 151-747,
KOREA, and\\
Physikalisches Institut, Universit\"{a}t Bonn, Nussallee 12,
D53115 Bonn, GERMANY
}

\begin{abstract}
It is argued that a phenomenologically viable
grand unification model from superstring is
$SU(3)^3$, the simplest gauge group among
the grand unifications of the electroweak hypercharge embedded
in semi-simple groups. We construct a realistic 4D
$SU(3)^3$ model with the GUT scale $\sin^2\theta_W^0=
\frac38$ in a $Z_3$ orbifold with Wilson line(s).
By two GUT scale vacuum expectation values, we obtain
a rank 4 supersymmetric standard model below the GUT scale,
and predict three more strange families.
\\
\vskip 0.5cm\noindent [Key words: $Z_{3}$ orbifold, 4D superstring,
trinification]
\end{abstract}


\maketitle
\newcommand{\bea}{\begin{eqnarray}}
\newcommand{\eea}{\end{eqnarray}}
\def\beq{\begin{equation}}
\def\eeq{\end{equation}}

\def\one{\bf 1}
\def\two{\bf 2}
\def\five{\bf 5}
\def\ten{\bf 10}
\def\tenb{\overline{\bf 10}}
\def\fiveb{\overline{\bf 5}}
\def\threeb{{\bf\overline{3}}}
\def\three{{\bf 3}}
\def\fb{{\overline{F}\,}}
\def\hb{{\overline{h}}}
\def\Hb{{\overline{H}\,}}

\def\slash#1{#1\!\!\!\!\!\!/}
\def\hf{\frac12}

\def\A{{\cal A}}
\def\Q{{\cal Q}}

\newcommand{\debug}{\emph{!!! CHECK !!!}}

\newcommand{\dd}{\mathrm{d}\,}
\newcommand{\Tr}{\mathrm{Tr}}
\newcommand{\drep}[2]{(\mathbf{#1},\mathbf{#2})}

\newpage

\def\sw{$\sin^2\theta_W$ }

\def\usw{$\sin^2\theta_W^0$ }

\subsection{Introduction and motivation}

Supersymmetric standard models(SSM), if proven
experimentally, need a theoretical explanation
of why they become the effective theory below the
Planck scale $M_P\simeq 2.44\times 10^{18}$ GeV.
A most probable scenario is that they result from
compactifications of superstring models preserving one
supersymmetry $N=1$. The effective 4D $N=1$ field theory
models were extensively considered in this regard in the
Calabi-Yau compactifications\cite{candelas} and orbifold
compactifications\cite{dhvw,inq}. Furthermore, the
standard-like models initiated more than 15 years ago
opened up the search for SSM
directly from superstring\cite{iknq}.

The initial standard-like
models $SU(3)\times SU(2)\times U(1)^n$ were very attractive,
in realizing the standard model(SM) gauge group and reasonable
matter spectrum\cite{iknq,imnq,others}, with possible
desirable physics on the strong CP problem\cite{cp}
and cosmology with a hidden world\cite{hidden}. Furthermore,
the doublet-triplet splitting has been realized in
some standard-like models\cite{iknq}.
However, these standard-like
models failed because they generally do not predict
correct weak mixing angle \usw\ at the string scale\cite{ibanez}.
To predict the observed coupling constants at the
electroweak scale successfully at least in $\sim 2.2\sigma$
level, the \usw\ at the unification scale $\sim (2-3)\times
10^{16}$ GeV is required to be $\simeq\frac38$.
The reason is very simple. In these standard-like models,
the electroweak hypercharge group $U(1)_Y$ is one
combination out of $n$ $U(1)$'s. Thus, the singlet
representations of the standard-like gauge group, not
belonging to the family structure of the fifteen(or sixteen
if we include a heavy Majorana neutrino),
can have nonvanishing $U(1)_Y$ charges, which lowers
the string scale weak mixing angle from the
needed value of $\frac38$, because the string scale
weak mixing angle \usw\ is expressed if we assume
$\alpha_2^0=\alpha_1^0$ at the string scale,
\begin{equation}\label{sinw}
\sin^2\theta_W^0=\frac{Tr\ T_3^2}{Tr\ Q_{em}^2}.
\end{equation}
This {\it sin$^2\theta_W$ problem} can be resolved if the
standard model
gauge group is unified in a simple group GUT, for
example $SU(5)$, where $U(1)_Y$ is a subgroup of the GUT
group. Then, the electroweak hypercharge generater is an
$SU(5)$ generator. Namely, $SU(5)$ singlets do not carry
nonvanishing electroweak hypercharges and we conclude that
the string scale \usw is $\frac38$.
To obtain a supersymmetric standard model in 4D, $SU(5)$
must be broken by a VEV of an adjoint Higgs field(${\bf 24}_H$).
However, it is impossible to obtain an adjoint matter
field at the level 1, i.e. $k=1$.
\footnote{At higher level $k>1$,
it was shown that the adjoint representation
{\bf 45} of $SO(10)$ can be
obtained\cite{tye}.}
If simplicity is any guidance to the truth of
nature, one must break the GUT group without an
adjoint matter representation. This leads us to GUT
groups with a $U(1)$ factor, notably $SU(5)\times U(1)$ which
is now called flipped $SU(5)$. The flipped $SU(5)$ is
an interesting rearrangement of a singlet field and
fifteen chiral fields
of $SU(5)$\cite{su51}. The symmetry breaking of the
flipped $SU(5)$ is particularly interesting in supersymmetric
flipped $SU(5)$\cite{susy51}. In this regards, the string
compactifications toward flipped $SU(5)$ is very interesting,
since breaking of $SU(5)\times U(1)$ down to the standard
model gauge group can be achieved without an adjoint Higgs
representation\cite{susy51}. Indeed, the fermionic
construction of 4D flipped $SU(5)$ was obtained
already fifteen years ago\cite{string51}. As shown in many
subsequent papers, the flipped $SU(5)$ has many
phenomenologically interesting features\cite{string51}.

However, the flipped $SU(5)$ generally fails in the
aforementioned \sw problem. The reason is the following.
The flipped $SU(5)$ needs three $SU(5)$ singlet representations
which carry +1 unit of the electric charge
for the three singlet charged leptons of SSM.
This implies, $SU(5)$ singlets can carry electromagnetic
charges, or the electroweak hypercharge $Y$. Since there appear
numerous $SU(5)$ singlets from string compactification,
the charged singlets generally reduce dramatically
\usw\ from the needed value $\frac38$,
 viz. (\ref{sinw}).

In the orbifold construction, this \sw problem
has been really serious. In the literature,
one can find many models with $SU(5)\times U(1)$
groups\cite{kanazawata}, and even it was
claimed that there are flipped $SU(5)$'s\cite{kanazawa},
but as shown above these models ignored
the \sw problem. However, one may argue that
even if the flipped $SU(5)$ contains a $U(1)$ factor,
the \sw problem goes away if the representations are
embeddable in $SO(10)$. In this case,
the $U(1)_Y$ generator belongs to $SO(10)$ and hence
$SO(10)$ singlets do not carry the $U(1)_Y$ charge.
Then, the singlets of the flipped $SU(5)$ carry
only the needed electroweak hypercharges of the flipped
$SU(5)$, and hence the string scale \usw\ is $\frac38$.
However, this scenario is not realized generally in orbifold
compactifications, which can be easily understood by
remembering that orbifolds generally choose
only part of the original complete representation.
In fact, this property is the root for the solutions of the
doublet-triplet splitting problem in the 4D orbifold
compactifications\cite{iknq}.

However, if it happens that the extra fields beyond
the complete multiplets conspire to contribute to
$Tr T_3^2$ and $Tr Q_{em}^2$ in the ratio 3/8, then we can
obtain 3/8 as the string scale value of \usw.
Therefore, the above argument is not a no-go theorem.
It may be extremely difficult however, if not impossible,
to find such a model with the
electroweak hypercharge leaking to $U(1)$ at the GUT scale.

Before considering our 4D string model, let us
comment on the recent field theoretic orbifold breaking
of grand unification group with extra
dimensions\cite{kawamura}. One interesting feature here has been
family unification groups with $SO(2n)$ with $n\ge 7$
\cite{fam}.
In these extra-dimensional field theories,
it is possible to allow fixed point fields as far as
there are no anomalies, and hence it is not much achieved
in the prediction of the matter representations
at the orbifold fixed points. In this
context, 6D string theoretic models were considered
as an intermediate step toward a final 4D string theory
construction\cite{ck}. In this paper, however, we attempt
to obtain a more ambitious 4D model.

\subsection{$Z_3$ orbifold with Wilson line}

\def\tri{$SU(3)^3$}

In 4D, if a GUT group containing a $U(1)$, as in the
$SU(5)\times U(1)$, is difficult to obtain,
the next simple
GUT groups to try are semi-simple groups. Therefore, we
propose {\it grand unified theories with
the hypercharge embedded in a semi-simple group
with no adjoint representation needed}(HESSNA)
as possible 4D string models toward a
realistic SSM. For a realistic
4D superstring model, we must require that the factor groups
of the HESSNA can be broken to SSM without an adjoint
representation. In this regard, note that the Pati-Salam
GUT group $SU(4)\times SU(2)_L\times SU(2)_R$ is
not a HESSNA because it has the same problem as that
in the $SU(5)$ model: one needs an adjoint representation.
Therefore, the simplest HESSNA is \tri.
The next simple HESSNA is $SU(3)\times SU(3)\times SU(4)$.
If we find a realistic HESSNA, then it is a simple matter
to find a SSM from this HESSNA, as the
SU(5) model leads to the SM.

In the HESSNA also, the orbifold compactification is
very much chiral, and may be too much chiral. But here at
least it is easy to study the electroweak
hypercharge concretely in a few steps.

At the phenomenological level, the group \tri\ has been
extensively considered\cite{trini}.
Our objective in this paper is to realize a string
theory \tri. If we obtain such a model, it can be considered
as a realistic superstring GUT.

We expect that one family in the \tri\ HESSNA is composed
of 27 chiral fields,
\begin{equation}\label{family1}
{\bf (\bar 3,3,1)+(1, \bar 3, \bar 3)+(3, 1, 3)}
\end{equation}
under \tri\ group. It can be embeddable in {\bf 27} of
$E_6$. Suppose, we assign the electroweak hypercharge
in $E_6$ such that the two neutral members in {\bf 27}
appear in the $SO(10)$ singlet and $SU(5)$ {\bf 10},
namely as in the flipped $SU(5)$ subgroup.
If we do that in $E_6$, $E_6$ is completely broken
down to the SM. Similarly,
two neutral members in the Higgs representation,
transforming like (\ref{family1}), are given large
HESSNA vacuum expectation values and a SSM can be obtained.

Only two possible \tri\ groups can be found in the
extensive tables of $Z_N$ orbifold models\cite{kanazawata}.
They appear in $Z_{12}$ orbifold models. However, the
fermionic spectrums of these $Z_{12}$ compactifications are
not the one required in (\ref{family1}).
This leads us to consider orbifold models with Wilson
lines.\footnote{
Since there does not exists
a complete table for $Z_M\times Z_N$ orbifolds,
we are not sure whether
HESSNA is possible for $Z_M\times Z_N$.}
In a separate publication, we tabulate
$Z_3$ orbifold models with one Wilson line\cite{chk}.

In the remainder of this paper, we present
a \tri\ model in a $Z_3$ orbifold compactification
with one Wilson line. Let us denote the $Z_3$ shift vector
as $v$ and the Wilson line as $a_1$. These must satisfy the
conditions for the shift vectors,
\begin{eqnarray}
v^2&=&\frac23\cdot({\rm integer}),\ \
a_1^2=\frac23\cdot({\rm integer}),\nonumber \\
(v_I)^2&=&
\frac29\cdot({\rm integer})\ {\rm for\ }I=\{1,2,\cdots, 8\}\
{\rm or\ }\{9,10,\cdots,16\},\\
a_{1I}^2&=&
\frac29\cdot({\rm integer})\ {\rm for\ }I=\{1,2,\cdots, 8\}\
{\rm or\ }\{9,10,\cdots,16\}\nonumber.
\end{eqnarray}
The modular invariance condition requires in addition,
\begin{eqnarray}
&3\ v\cdot a_i = {\rm (integer)\ for\ }
(i=1,3,5),\nonumber\\
&3\ a_i\cdot a_j
={\rm (integer)}\ {\rm for\ } i\ne j.
\end{eqnarray}
The notation is the same as those discussed in\cite{notation}.
For an \tri\ gauge group,
we choose the following shift vector and a Wilson line,
\begin{eqnarray}\label{model}
&v=\left( 0~0~0~0~0~
\frac13~ \frac13~ \frac23
\right)
\left(
0~0~0~ 0~0~0~0~0
\right)\nonumber\\
&a_1=\left(
\frac13~ \frac13~ \frac13~ 0~0~
\frac13~ \frac13~ \frac53
\right)\left(
0~0~0~0~0~0~0~0
\right)
\end{eqnarray}

\subsection{Untwisted sector}

\noindent {\bf Gauge group :} From the mass shell condition
$ {m^2 \over 4} = {p^2 \over 2} - 1$, we find the massless
spectrum in the untwisted sector. For the
gauge bosons, the $p^2=2$ root vectors, satisfying
$p\cdot v=0$ and $p\cdot a_1=0$ mod integer, are
the nonvanishing roots. These are presented for the
first $E_8$ subgroup in Table \ref{tab1}.
The second $E_8^\prime$ gauge group is not broken.

\begin{center}
\begin{table}
\caption{\label{tab1}\it Root vectors $p_I$ in untwisted sector satisfying
$p\cdot v=0$ and $p\cdot a_1=0$. The underlined entries
allow permutations. The $+$ and $-$ in the spinor part
denote $\frac12$ and $-\frac12$, respectively. $I, V,$
and $U$ spin directions of $SU(3)$'s are also shown.}
\vskip 0.3cm
\begin{center}
\begin{tabular}{|c|c|c|}
\hline
vector & number of states  & gauge group \\
\hline
$ (\underline{1~ -1~~\ 0}~~\ 0~~\ 0~~\ 0~~\ 0~~\ 0) \ \ $ & 6 & $SU(3)_1$\\
\hline
$ (0~~\ 0~~\ 0\ ~~ \ 1~~ \ 1~~\ 0~~\ 0~~\ 0)_{I_+} $ & 1 &\\
$ ( 0~~\ 0~~\ 0~ -1 -1~\ 0~~\ 0~~\ 0)_{I_-} $ & 1 &\\
$ (+~+~+~+~+~-~-~+)_{V_+} $ & 1 & \\
$ (-~-~-~-~-~+~+~-)_{V_-} $ & 1 & $SU(3)_2$\\
$ (+~+~+~-~-~-~-~+)_{U_+} $ & 1 & \\
$ (-~-~-~+~+~+~+~-)_{U_-} $ & 1 & \\
\hline
$ (0~~\ 0~~\ 0~~\ 1~-1~~\ 0~~\ 0~~\ 0)_{I_+} $ & 1 &\\
$ (0~~\ 0~~\ 0~-1~~\ 1~~\ 0~~\ 0~~\ 0)_{I_-} $ & 1 &\\
$ (+~+~+~+~-~+~+~-)_{V_+} $ & 1 & \\
$ (-~-~-~-~+~-~-~+)_{V_-} $ & 1 & $SU(3)_3$\\
$ (+~+~+~-~+~+~+~-)_{U_+} $ & 1 & \\
$ (-~-~-~+~-~-~-~+)_{U_-} $ & 1 & \\
\hline
$ (0~~\ 0~~\ 0~~\ 0~~\ 0~~\underline{\ 1~-1}~\ 0)_{I_{\pm}} $ & 2 & \\
$ (0~~\ 0~~\ 0~~\ 0~~\ 0~~\ 0 -1 -1)_{V_+} $ & 1 & \\
$ (0~~\ 0~~\ 0~~\ 0~~\ 0~~\ 0~~\ \ 1~~\ \  1)_{V_-} $ & 1 & $SU(3)_4$\\
$ (0~~\ 0~~\ 0~~\ 0~~\ 0 -1~~\ 0 -1)_{U_+} $ & 1 & \\
$ (0~~\ 0~~\ 0~~\ 0~~\ 0~~\ 1~~\ 0~~\ \ 1)_{U_-} $ & 1 & \\
\hline
\end{tabular}
\end{center}
\end{table}
\end{center}
\vskip 0.2cm

In Table \ref{tab1}, we use the convention that the underlined entries
allow permutations. There are 6 winding states in the first row and
adding two oscillators we have the 8 roots for the first $SU(3)_1$.
Similarly, we obtain the rest $SU(3)$'s.
Thus, we obtain the gauge group \tri $\subset SU(3)^4$
with the corresponding
nonvanishing root vectors explicitly shown. Note in passing
that there is no $U(1)$ subgroup, which means that there
is no anomalous $U(1)$ gauge group with the above orbifold.
Thus, it is possible to realize the model-independent
axion as a quintessential axion\cite{kn03}.
\\

\noindent{\bf Matter from the untwisted sector :} The matter fields
from the untwisted sector satisfy the condition
\begin{equation}
p^2=2,\ \ p\cdot v=\frac23\ {\rm mod\ integer},\ \ p\cdot a_i=0
\ {\rm mod\ integer}.
\end{equation}
In Table \ref{tab2}, we present the root vectors satisfying these.

\begin{center}
\begin{table}
\caption{\label{tab2}
\it Root vectors $p_I$ in untwisted sector satisfying
$p\cdot v=\frac23$ and $p\cdot a_1=0$. The underlined entries
allow permutations. The notations are the same as in Table I,
except that} [\ ] {\it implies even numbers of sign flips.
In the last column, we reverseed the chirality to compare directly
with the twisted sectors.}
\vskip 0.2cm
\begin{center}
\begin{tabular}{|c|c|c|}
\hline
 \raisebox{-1.8ex}[0pt][0pt]{sector}
            &
         \multicolumn{2}{|c|}{From $E_8$ roots}
 \\
 \cline{2-3}
 & \multicolumn{1}{|c|}{$E_8$ root} & \multicolumn{1}{|c|}{$SU(3)^4$}
 \\
\hline
 & $ (\underline{~1~~\ 0~~\ 0}~~\ 0~~\ 0~\underline{-1~\ 0}~~\ 0~) $  & \\
\raisebox{-1.8ex}[0pt][0pt]{UT} & $ (\underline{~1~~\ 0~~\ 0}~~\ 0~~\ 0~~\ 0~~\ 0~~\ 1~) $  &\\
 & $ (\underline{+~-~~-}~[+~~+]~~\underline{+~~-}~~-) $
& \raisebox{1.8ex}[0pt][0pt]{ 3${\bf (\bar 3, 3,1,\bar 3)}$}\\
 & $ (\underline{+~-~~-}~[+~~+]~~+~+~+) $
&  \\
\hline
\end{tabular}
\end{center}
\end{table}
\end{center}
\vskip 0.2cm

\subsection{Matter from the twisted sectors}

In $Z_3$ orbifolds, there are three fixed point
on a 2-torus. Since we compactify six internal
spaces via three 2-tori, there are 27 fixed points.
These 27 fixed points look the same in every aspect
if we do not introduce Wilson lines. If we allow
the possibility to wrap the 2-torus by a Wilson line,
then three fixed points on the torus can be
distinguished by the gauge fields going around the
torus. There are two directions to wrap the torus, but
the modular invariance requires that
they must be the same, i.e. $a_1=a_2$. Similarly, if we
wrap more tori, we have $a_3=a_4$
and $a_5=a_6$\cite{inq}. Thus, we can consider at most three
independent Wilson lines, $a_1,a_3$, and $a_5$. In
this paper, we considered the simplest Wilson line, i.e.
$a_1\ne 0$, and $a_3=a_5=0$. So the 27 fixed points
are grouped into three classes: 9 trivial fixed
points around which there is no Wilson line($v$),
9 positively wraped fixed points($v+a_1$),
and 9 negatively wraped fixed points($v-a_1$),
which are denoted as T0, T1, and T2 twisted
sectors, respectively.

In our model, the massless matter
fields from the twisted sectors satisfy
$(p+\tilde v)^2=\frac23,\frac43$, where
$\tilde v=v, v+a_1, v-a_1$, for T0, T1, and T2, respectively.
Of course, the weights we present
survive the GSO-like projection.
For the vectors corresponding to $\frac43$ the multiplicity
is 9 as described above, and for the vectors
corresponding to $\frac23$ the
multiplicity is 27 because of the three oscillator modes
in this case.

In general, the matter fields from the twisted sectors make the
theory extremely chiral which was the reason that we have not
obtained yet any realistic SSM or flipped $SU(5)$ model
from orbifold compactification of the heterotic string. Since it
is very chiral, there is a chance that the spectrum
(\ref{family1}) can appear through orbifolding.

In Tables III, IV, and V, we list the massless
spectrum from the twisted sectors. But note
that the chirality of the
twisted sector in the $Z_3$
orbifold is the opposite of the chirality of the
untwisted sector matter fields.

\begin{center}
\begin{table}
\caption{\label{tab3}
\it Root vectors $p_I$ in the T0 twisted sector
satisfying $p\cdot \tilde v=\frac23,\frac43$. The notations
are the same as in Table II.}
\vskip 0.3cm
\begin{center}
\begin{tabular}{|c|c|c|}
\hline
 \raisebox{-1.8ex}[0pt][0pt]{sector}  &   \multicolumn{2}{|c|}{
 Weights} \\
 \cline{2-3}
 & \multicolumn{1}{|c|}{vector} & \multicolumn{1}{|c|}{$SU(3)^4$}
 \\
\hline
 & $ (0~~0~~0~~0~~0~~ \ \ 0~~ \ \ 0~~ \ \ 0) $  & \\
\raisebox{-9.2ex}[0pt][0pt] &
$ (0~~0~~0~~0~~0~-1~~ \ 0~-1) $  & 27${\bf (1,1,1,
 3)}$\\ & $ (0~~0~~0~~0~~0~~ \ 0~-1~-1) $  & \\
\cline{2-3}
 & $ (\underline{-1~ 0~\ 0}~~ \ 0~~\ 0~~\ 0~~ \ 0~ -1) $  &\\
 & $ (\underline{+~+~-}~[+~+]~-~-~-) $
&\raisebox{1.6ex}[0pt][0pt]{ 9${\bf (\bar 3,3,1,1)}$}\\
\cline{2-3}
T0 & $ (\underline{1~~\ 0~~\ 0}~~\ 0~~\ 0~~\ 0~~\ 0 -1) $  &\\
 & $ (\underline{+~-~-}~[+~-]~-~-~-) $
&\raisebox{1.6ex}[0pt][0pt]{ 9${\bf (3,1,\bar 3,1)}$}\\
\cline{2-3}
 & $ (0~~\ 0~~\ 0~~\ 0~~\ 0 -1 -1~\ 0) $  &\\
 & $ (0~~\ 0~~\ 0~~\underline{1~~\ 0}~~\ 0~~\ 0 -1) $  &\\
 & $ (0~~\ 0~~\ 0~\underline{-1~~\ 0}~~\ 0~~\ 0 -1) $
& 9${\bf (1,\bar 3, 3,1)}$\\
 & $ (+~+~+~[+~-]~-~-~-) $  &\\
 & $ (-~-~-~[+~+]~-~-~-) $  & \\
\hline
\end{tabular}
\end{center}
\end{table}
\end{center}

\begin{center}
\begin{table}
\caption{\label{tab4}
\it Root vectors $p_I$ in the T1 twisted sector.}
\vskip 0.3cm
\begin{center}
\begin{tabular}{|c|c|c|}
\hline
 \raisebox{-1.8ex}[0pt][0pt]{sector}  &   \multicolumn{2}{|c|}{
 Weights} \\
 \cline{2-3}
 & \multicolumn{1}{|c|}{vector} & \multicolumn{1}{|c|}{$SU(3)^4$}
 \\
\hline
 & $ (0~~0~~0~~0~~0~-1~-1~-2) $  & \\
 & $ (-~-~-~[+~~+]~-~-~\frac{-5}{2}) $
& \raisebox{1.6ex}[0pt][0pt]{ 27${\bf (1,\bar 3,1,1)}$}\\
\cline{2-3}
 & $ (0~~\ 0~~\ 0~~\ 0~~\ 0~\underline{-1~~0}~-3) $& \\
 & $ (0~~\ 0~~\ 0~~\ 0~~\ 0~~\ 0~~\ 0 -2) $  & \\
T1 & $ (-~-~-~[+~-]~\underline{-~~\frac{-3}{2}}~
 \frac{-5}{2}) $  & \raisebox{1.6ex}[0pt][0pt]{ 9
 ${\bf (1,1,\bar 3, 3)}$}\\
 & $ (-~-~-~[+~-]~-~-~\frac{-3}{2}) $  & \\
\cline{2-3}
 & $ (\underline{-1~~0~~0}~~0~~0~-1~-1~-3) $  &\\
 & $ (\underline{-1~~\ 0~~\ 0}~~\ 0~~\ 0~~\underline{-1~~0}
 ~-2) $  & \raisebox{1.6ex}[0pt][0pt]{ 9
${\bf (\bar 3,1,1,\bar 3)}$}\\
\cline{2-3}
 & $ (\underline{+~-~-}~~[+~-~]~-~-~\frac{-5}{2}) $
& \raisebox{-1.6ex}[0pt][0pt]{ $9{\bf ( 3,1,3,1)}$} \\
 & $ (\underline{-1~-1~~0}~~0~~0 -1 -1 -2) $&  \\
\hline
\end{tabular}
\end{center}
\end{table}
\end{center}

\begin{center}
\begin{table}
\caption{\label{tab5}
\it Root vectors $p_I$ in the T2 twisted sector.}
\begin{center}
\vskip 0.3cm
\begin{tabular}{|c|c|c|}
\hline
 \raisebox{-1.8ex}[0pt][0pt]{sector}  &   \multicolumn{2}{|c|}{
 Weights} \\
 \cline{2-3}
 & \multicolumn{1}{|c|}{vector} & \multicolumn{1}{|c|}{$SU(3)^4$}
 \\
\hline
& $ (\underline{~1~~\ 0~~\ 0}~~\ 0~~\ 0~~\ 0~~\ 0 -1) $
& 27${\bf (3,1,1,1)}$\\
\cline{2-3}
 & $ (+~+~+~[+~-]~+~-~+) $  &\\
 & $ (0~~\ 0~~\ 0~~\ 0~~\ 0~~\ 0 -1~~\ 1) $  & \\
 & $ (+~+~+~[+~-]~-~+~+) $  & \\
 & $ (0~~\ 0~~\ 0~~\ 0~~\ 0 -1~~\ 0~~\ 1) $
& \raisebox{1.6ex}[0pt][0pt]{ 9${\bf (1,1,3,3)}$}\\
 & $ (+~+~+~[+~-]~+~+~\frac{+3}{2}) $ & \\
 & $ (0~~\ 0~~\ 0~~\ 0~~\ 0~~\ 0~~\ 0~~\ 2) $  & \\
\cline{2-3}
 & $ (1~~\ 1~~\ 1~~\ 0~~\ 0~~\ 0~~\ 0~~\ 1) $  & \\
 & $ (+~+~+~[+~+]~+~+~+) $  &    \\
T2 & $ (+~+~+~+~-~-~-~\frac{+3}{2}) $  & \\
 & $ (0~~\ 0~~\ 0~~\ 1~~\ 0~~\ 0~~\ 0~~\ 1) $
& \raisebox{-1.6ex}[0pt][0pt]{ 9 ${\bf (1,\bar 3,\bar 3,1)}$}\\
 & $ (0~~\ 0~~\ 0~~\ 0 -1~~\ 0~~\ 0~~\ 1) $  & \\
 & $ (+~+~+~-~+~-~-~\frac{+3}{2}) $  & \\
 & $ (0~~\ 0~~\ 0~~\ 0~~\ 1~~\ 0~~\ 0~~\ 1) $  & \\
 & $ (0~~\ 0~~\ 0 -1~~\ 0~~\ 0~~\ 0~~\ 1) $  & \\
\cline{2-3}
 & $ (+~+~+~[+~+]~-~-~+) $  &    \\
 & $ (0~~\ 0~~\ 0~~\ 0~~\ 0~~\ 0~~\ 0~~\ 0) $  & \\
 & $ (+~+~+~[+~+]~+~-~\frac{+3}{2}) $  & \\
 & $ (0~~\ 0~~\ 0~~\ 0~~\ 0~~\ 1~~\ 0~~\ 1) $
& \raisebox{1.6ex}[0pt][0pt]{ 9 ${\bf (1,3,1,\bar 3)}$}\\
 & $ (+~+~+~[+~+]~-~+~\frac{+3}{2}) $  & \\
 & $ (0~~\ 0~~\ 0~~\ 0~~\ 0~~\ 0~~\ 1~~\ 1) $  & \\
\hline
\end{tabular}
\end{center}
\end{table}
\end{center}

\subsection{Electroweak hypercharge}

In the \tri\ GUT, the color factor should not carry the
electroweak hypercharge. To break $SU(3)^4$ gauge group
down to \tri\, another $SU(3)$ should not carry the
hypercharge. Let us break $SU(3)_4$ completely by two
independent vacuum expectation values of ({\bf 1,1,1,3}).
Thus the \tri\ group is $SU(3)_1\times SU(3)_2\times
SU(3)_3$. We identify $SU(3)_2$ as the group containing
the $W^\pm$ bosons and $SU(3)_3$ as QCD.
Under the $SU(3)_1\times SU(3)_2\times SU(3)_3$,
we obtain the following chiral fermions,
\begin{eqnarray}
&9\left[{\bf (\bar 3,3,1)+(1,\bar 3,\bar 3)+(3,1,3)}\right]_a
                                                 \nonumber\\
&+9\left[{\bf (\bar 3,3,1)+(1,\bar 3,3)+(3,1,\bar 3)}\right]_b
+\cdots\label{trist}
\end{eqnarray}
where $\cdots$ represents 27 multiplets of the
vectorlike combination  {\bf (3,1,1)}
$+{\bf (\bar 3,1,1)}$ ${\bf +(1,3,1)}$  $+{\bf (1,\bar 3,1)}$
${\bf +(1,1,3)}$  $+{\bf (1,1,\bar 3)}$ +3${\bf (1,1,1)}$.
Eq. (\ref{trist}) realizes
the representation given in (\ref{family1}).

The hypercharge($\equiv$ electroweak hypercharge) $Y$
is a combination of generators of $SU(3)_1$ and $SU(3)_2$,
\begin{equation}
Y=-\frac12 (-2I_1 + Y_1 + Y_2)
\end{equation}
where $I_1$ is the third component $(T_3)_1$ of the isospin
generators
of the group $SU(3)_1$, and $Y_i$ is the $SU(3)_i (i=1,2)$
hypercharge $\frac{2}{\sqrt{3}}(T_8)_i$.
The eigenvalues of $I$ and $Y$ are $\{\frac12,
-\frac12,0\}$ and $\{\frac13,\frac13,-\frac23\}$, respectively.
One can easily check that the model presented
in (\ref{trist}) gives \usw$=\frac{3}{8}$, thus
solving the string \sw problem.
Since the hypercharge
$U(1)_Y$ does not leak to $SU(3)_3$(=QCD), in counting
the eigenvalues of the electroweak $(T_3)_2^2$ and
$Q_{em}^2$, the contributions from
the first and the second [\ ] brackets of
Eq. (\ref{trist}) are exactly the same.
We also checked that the vectorlike representation
contributes in the same ratio. There
unfamiliar particles such as lepton doublets with
$Y=\pm\frac16$ appear, but they form a vectorlike representation,
are removed at the GUT scale and do
not alter \usw. This miraculous prediction of
\usw\ is based on the fact that everything appears in the
multiples of 3. The model given in
Eq. (\ref{trist}) gives 9 families. But note that
there appear additional 9 families with the opposite
colors. By adding more Wilson line(s) in the
hidden sector $E_8^\prime$
part, the family number can be easily reduced to 3,
not spoiling our precious spectrum obtained in (\ref{family1}).
Below, we comment on six family models obtained by adding
more Wilson line(s) at $E_8^\prime$.

The spectrum (\ref{trist}) has two villages, each having three
families. The family mixing is allowed inside the village
but is forbidden between different villages, predicting two
CP phases, one in each village. To explain the three light
families, the members of the strange village are required to
be heavy at the electroweak scale.
With 6 families, the QCD coupling is not asymptotically
free, but still perturbatively unifiable at the GUT scale.
The rank--6 \tri\ is directly broken down to the rank--4 SSM
by two vacuum expectation values in ${\bf (\bar 3, 3,1)}$, i.e.
$\langle{\bf (1_{\bar 3},1_3,1) }\rangle
=$(GUT scale) and $\langle{\bf (2^{\downarrow}_{\bar 3},
1_3,1)}\rangle=$(GUT scale), where
${\bf 2^{\downarrow}_{\bar 3}}$
is the $(I_3)_1=-\frac12$ member in ${\bf\bar 3}$ of
$SU(3)_1$, etc.
With some hypotheses on removing a set of
vectorlike representations at the GUT scale, we obtain a
realistic SSM in the present orbifold compactification
with the help of the GUT scale VEV's. Here, we assume that
three SM singlets in our village are removed
at high energy scale, but the three singlets of the stranger
village are left light so that they can acquire Dirac masses
at the electroweak symmetry breaking scale. Namely, we will have
three light neutrinos.

The three strange village families presented in
(\ref{trist}) may be considered
as a drawback of the present construction. But remembering
the enormous difficulties during the last two decades
in obtaining a superstring derived
SSM, in view of a model like (\ref{trist})
we may envision a Planck scale string theory
verifiable through TeV--scale probing colliders.

\acknowledgments
I thank K.-S. Choi, K. Hwang, and H. P. Nilles
for helpful discussions. I thank the Physikalisches Institute
of Univ. of Bonn, and in particular Hans Peter Nilles,
for the hospitality during my stay for this work.
I also thank the Humboldt Foundation
for the award. This work is supported in part by the BK21
program of Ministry of Education, and the KOSEF Sundo Grant.

\end{document}